\documentclass[aps,prd,twocolumn,twoside,showpacs,showkeys]{revtex4}
\usepackage{amsmath,mathptmx}

\begin{document}
\bibliographystyle{apsrev} 

\title{Oscillations do not distinguish between massive and tachyonic
  neutrinos}

\author{Pawe{\l} Caban}
\email{P.Caban@merlin.fic.uni.lodz.pl}
\author{Jakub Rembieli{\'n}ski}
\email{J.Rembielinski@merlin.fic.uni.lodz.pl}
\author{Kordian A. Smoli{\'n}ski}
\email{K.A.Smolinski@merlin.fic.uni.lodz.pl}
\author{Zbigniew Walczak}
\email{Z.Walczak@merlin.fic.uni.lodz.pl}
\affiliation{Department of Theoretical Physics, University of {\L}{\'o}d{\'z},
  ul.~Pomorska~149/153,~90-236~{\L}{\'o}d{\'z}, Poland}
\date{\today}
\begin{abstract}
  It is shown that the hypothesis of tachyonic neutrinos leads to the
  same oscillations effect as if they were usual massive particles.
  Therefore, the experimental evidence of neutrino oscillations does
  not distinguish between massive and tachyonic neutrinos.
\end{abstract}
\maketitle

In the last two decades one of the most interesting problems in the
particle physics is the issue of neutrino masses \cite{gelmini95}.
The flavor oscillations \cite{altmann01,beuthe03} are the indirect
evidence for non-zero neutrino masses.  On the other hand, the
experiments devoted to direct measurement of the neutrino mass
permanently yield the negative mass squared for the electron neutrino
\cite{lobashev99,weinheimer99} and the muon neutrino
\cite{assamagan96} (for a review, see also \cite{hagiwara02}).  In
particular, the most sensitive neutrino mass measurement involving
electron antineutrino, is based on fitting the shape of beta spectrum.
An as yet not understood event excess near the spectrum endpoint can
be explained \cite{ciborowski97,ciborowski99} on the ground of the
hypothesis of the tachyonic neutrinos \cite{chodos85}.  The above
hypothesis can be also helpful in explanation of some effects in
cosmic ray spectrum \cite{ehrlich99a,ehrlich99b,ehrlich00}.  From the
theoretical point of view, this hypothesis can be formulated
consistently only in the framework of the absolute synchronization
scheme \cite{rembielinski97,caban99} which needs the notion of
preferred frame.

The aim of this report is to compare the flavor change effect for the
massive and tachyonic neutrinos.  Calculations will be performed in
the preferred frame, in which the metric tensor has the standard
Minkowskian form.

Let us consider three flavors of neutrinos.  We shall denote the
neutrino flavor states by
\begin{equation}
  \label{eq:1}
  |\nu_e\rangle = 
  \begin{pmatrix}
    1\\
    0\\
    0
  \end{pmatrix}
  \otimes |\vec{p}\rangle, \quad
  |\nu_\mu\rangle = 
  \begin{pmatrix}
    0\\
    1\\
    0
  \end{pmatrix}
  \otimes |\vec{p}\rangle, \quad
  |\nu_\tau\rangle = 
  \begin{pmatrix}
    0\\
    0\\
    1
  \end{pmatrix}
  \otimes |\vec{p}\rangle.
\end{equation}
The $|\vec{p}\rangle$ denotes the vector from the representation space of
the Poincar{\'e} group, corresponding to the momentum $\vec{p}$.

In the basis~\eqref{eq:1} the free Hamiltonian describing neutrinos
can be written as
\begin{equation}
  \label{eq:5}
  H = U^{\dag} H_0 U,
\end{equation}
where $U$ is a $3 \times 3$ unitary mixing matrix, acting in each subspace
of the Hilbert space determined by momentum $\vec{p}$, and the
Hamiltonian $H_0$ in the mass eigenstates basis is of the form
\begin{equation}
  \label{eq:4}
  H_0 =
  \begin{pmatrix}
    \sqrt{\vec{p}^2 \pm m_1^2} & 0 & 0 \\
    0 & \sqrt{\vec{p}^2 \pm m_2^2} & 0 \\
    0 & 0 & \sqrt{\vec{p}^2 \pm m_3^2}
  \end{pmatrix}
  .
\end{equation}
Here $m_j^2$ ($j = 1, 2, 3$) are the absolute values of the squares of
neutrino four-momenta and the choice of the sign corresponds to the
massive ($+$) and tachyonic ($-$) case, moreover, $|\vec{p}| > \max_j
m_j$ in the tachyonic case.

Following the standard procedure, we consider the time evolution of
the neutrino state.  Assume that the initial state $|\nu_i\rangle$ ($i = e, \mu,
\tau$) is an eigenstate of the fractional lepton number operator.  Thus,
after time $t$ the probability that we get the neutrino $\nu_k$ is given
by
\begin{equation}
  \label{eq:7}
  P_{\nu_i \to \nu_k}(t) 
  = \left|\langle\nu_k| U^{\dag} e^{-i t H_0} U |\nu_i\rangle\right|^2.
\end{equation}

Now, taking~\eqref{eq:7} in the limit $|\vec{p}| \gg m_j$, we can expand
the matrix elements of the Hamiltonian as follows
\begin{equation}
  \label{eq:12}
  \sqrt{\vec{p}^2 \pm m_i^2} \simeq p \pm \frac{m_i^2}{2 p}.
\end{equation}
Using~\eqref{eq:12} we obtain from~\eqref{eq:7}
\begin{equation}
  \label{eq:14}
  P_{\nu_i \to \nu_k}(t) = \left|\sum_{j=1}^3 u_{jk}^* u_{ji} e^{\mp\frac{i t}{2 p} m_j^2}\right|^2.
\end{equation}

On introducing the polar representation such that
\begin{subequations}
  \label{eq:16}
  \begin{align}
  u_{2k}^* u_{2i} u_{3k} u_{3i}^* &= r_1(i,k) e^{i \theta_1(i,k)},\\
  u_{3k}^* u_{3i} u_{1k} u_{1i}^* &= r_2(i,k) e^{i \theta_2(i,k)},\\
  u_{1k}^* u_{1i} u_{2k} u_{2i}^* &= r_3(i,k) e^{i \theta_3(i,k)},
\end{align}
\end{subequations}
and
\begin{equation}
  \label{eq:18}
  \omega_1 = \frac{m_2^2 - m_3^2}{2 p},\quad
  \omega_2 = \frac{m_3^2 - m_1^2}{2 p},\quad
  \omega_3 = \frac{m_1^2 - m_2^2}{2 p},
\end{equation}
we can write~\eqref{eq:14} in the following form:
\begin{equation}
  \label{eq:19}
  P_{\nu_i \to \nu_k}(t) 
  = \sum_{j=1}^3 |u_{jk}|^2 |u_{ji}|^2 + 2 \sum_{j=1}^3 r_j(i,k) \cos[\omega_j t \pm \theta_j(i,k)],
\end{equation}
where, as before, the signs $+$ and $-$ correspond to massive and
tachyonic neutrinos, respectively.  In particular, if $k = i$ then
$\theta_j(i,i) = 0$ ($j = 1, 2, 3$) and
\begin{equation}
  \label{eq:17}
  P_{\nu_i \to \nu_i}(t) = \sum_{j=1}^3 |u_{ji}|^4 + 2 \sum_{j=1}^3 r_j(i,i) \cos(\omega_j t),
\end{equation}
so this probability is the same for massive and tachyonic neutrinos.

We have shown that the only difference between the neutrino
oscillations in the massive and tachyonic case lies in the initial
phase of oscillations $\theta_j(i,k)$.  Moreover, the initial phases for
tachyonic case are obtained by taking the complex conjugations of the
elements of mixing matrix for the massive case (see
Eqs.~\eqref{eq:16}).  In the oscillation experiments we cannot decide
wheter we should take the mixing matrix $U$ or its complex conjugation
$U^*$ without taking into account the nature of the neutrinos.  More
precisely, we can explain the oscillations for massive or tachyonic
neutrinos simply by the different choice of the mixing matrix ($U$ or
$U^*$).  Therefore, the experimental evidence of neutrino oscillations
does not distinguish between massive and tachyonic neutrinos which,
together with other experiments
\cite{lobashev99,weinheimer99,assamagan96} leave the question of
tachyonic character of the neutrinos still open.

%\bibliography{tachyon,neutrino,pdg}

\end{document}